\documentclass[english]{elsarticle}
\usepackage[T1]{fontenc}
\usepackage[latin9]{inputenc}
\usepackage{amstext}
\usepackage{amssymb}
\usepackage{babel}
\def\be{\begin{equation}}
\def\ee{\end{equation}}
\def\pf{\textrm{pf}}
\journal{Computer Physics Communications}

\begin{document}

\begin{frontmatter}

\title{Numeric and symbolic evaluation of the pfaffian of general skew-symmetric
matrices}

\author[a]{C. Gonz\'alez-Ballestero}
\author[a]{L.M. Robledo\corref{author}}
\author[b]{G. F. Bertsch}

\cortext[author] {Corresponding author.\\\textit{E-mail address:} luis.robledo@uam.es}
\address[a]{Departamento de F\'\i sica Te\'orica, Universidad Aut\'onoma de Madrid, E-28049
Madrid, Spain}
\address[b]{Department of Physics and Institute for Nuclear Theory, University
of Washington, Seattle, WA 98195\textendash{}1560 USA}

\begin{abstract}
Evaluation of pfaffians arises in a number of physics applications, 
and for some of them a direct method is preferable to using the 
determinantal formula.  We discuss two methods for the numerical 
evaluation of pfaffians. The first is tridiagonalization based on 
Householder transformations. The main advantage of this method is 
its numerical stability that makes unnecessary the implementation of 
a pivoting strategy. The second method considered is based on 
Aitken's block diagonalization formula. It yields to a kind of LU 
(similar to Cholesky's factorization) decomposition (under congruence) of arbitrary 
skew-symmetric matrices that is well suited both for the numeric and 
symbolic evaluations of the pfaffian. Fortran subroutines (FORTRAN 
77 and 90) implementing both methods are given. We also provide 
simple implementations in Python and Mathematica for purpose of testing, or for 
exploratory studies of methods that make use of pfaffians.
\end{abstract}
\end{frontmatter}


{\bf PROGRAM SUMMARY}

\begin{small}
\noindent
{\em Manuscript Title:}   Numeric and symbolic evaluation of the pfaffian of general skew-symmetric matrices                                    \\
{\em Authors:} C. Gonzalez-Ballestero, L.M.Robledo and G.F. Bertsch      \\
{\em Program Title:} Pfaffian                                \\
{\em Journal Reference:}                                      \\
{\em Catalogue identifier:}                                   \\
{\em Licensing provisions:}                                   \\
{\em Programming language:}  Fortran 77 and 90                \\
{\em Computer:}                                               \\
{\em Operating system:}                                       \\
{\em RAM:} bytes                                              \\
{\em Number of processors used:}                              \\
{\em Supplementary material:}                                 
{\em Classification:} 4.8  Linear Equations and Matrices      \\
{\em External routines/libraries:}  BLAS                          \\
{\em Subprograms used:}                                       \\
{\em Catalogue identifier of previous version:}*              \\
{\em Journal reference of previous version:}*                  \\
{\em Does the new version supersede the previous version?:}*   \\
{\em Nature of problem:} Evaluation of the Pfaffian of a skew symmetric matrix. \\
 Evaluation of pfaffians arises in a number of physics applications 
 involving fermionic mean field wave functions and their overlaps.
   \\
{\em Solution method:} Householder tridiagonalization. 
Aitken's block diagonalization formula.\\
   \\
{\em Reasons for the new version:}*\\
   \\
{\em Summary of revisions:}*\\
   \\
{\em Restrictions:}\\
   \\
{\em Unusual features:}\\
   \\
{\em Additional comments:} Python and Mathematica implementations are
provided in the main body of the paper\\
   \\
{\em Running time:} Depends on the size of the matrices. For matrices
with 100 rows and columns a few miliseconds are required. \\
   \\

\end{small}

\section{Introduction}

In a number of fields in physics, the formal equations derived from 
the theory make use of the pfaffian of some skew-symmetric matrix 
appearing in the theory.  For example, the pfaffian arises in the 
treatment of electronic structure with quantum Monte Carlo methods \cite{Bajdich.08},
the description of two-dimensional Ising spin glasses 
\cite{Thomas.09}, and the evaluation of entropy and its relation to 
entanglement \cite{Stephan.09}.  Pfaffians occur naturally in field 
theory and nuclear physics in formalisms based on fermionic 
coherent states \cite{Berezin.66,Ohnuki.78,Klauder.85,Lang.93}.  A 
recent application is to the overlap of  two Hartree Fock Bogoliubov 
(HFB) product wave functions \cite{Robledo.09}, needed for nuclear 
structure theory.   While there is a simple formula for the pfaffian 
of a skew-symmetric matrix $M$ in terms of the determinant,
\be
\pf(A) = \sqrt{ {\rm det}(A)}
\ee
the so-called {}``sign problem of the overlap''
\cite{Neeargard.83} associated with the square root motivates the use
of numerical algorithms that evaluate it directly.  The most straightforward
method, the rule of {}``expanding in minors''
\cite{Caianello}, has bad scaling with the size of the matrix and is
prohibitive for large matrices.  In this paper we discuss two alternative
methods that have the same scaling property as the normal $N^3$ algorithms for
the determinant. The methods are implemented in the
FORTRAN 77 and 90 subroutines provided in the accompanying program 
library. We also comment on the practical implementation of the two methods 
in Mathematica and in the Python programming language.

\section{Evaluation of the Pfaffian}

The Pfaffian $\pf(A)$ is reduced to a simple form that is easily evaluated by
making repeated use of transformation formula given in \ref{sec:AppA}, 
\be
\pf( B^t A B) = {\rm det(B)} \pf(A).
\ee
In order to perform the numerical evaluation of the Pfaffian of a
complex skew-symmetric matrix $A$ we reduce the skew-symmetric
matrix to a tridiagonal form $A_{TR}$ by using unitary matrices $U$.
Once it is in this form, the evaluation
of the pfaffian is straightforward (see below).

\subsection{Reduction to tridiagonal form by mean of Householder transformations}

In this method, we will use the well-known Householder transformations \cite{Golub.96}
to reduce $A$ to tridiagonal form.  We present it in some detail because
the generalization to the complex number field is not entirely trivial. 

Complex Householder transformations have the form 
\begin{equation}
P=\mathbb{I}-2\frac{u\otimes u^{+}}{\left|u\right|^{2}}\label{eq:Householder}
\end{equation}
where $u$ is an arbitrary complex complex row vector $u=(u_{1},u_{2},\ldots,u_{N})$
and $\left(u\otimes u^{+}\right)_{ij}=u_{i}u_{j}^{*}$. The vector $u$ must 
be chosen to zero all the elements
of a vector $x$ except a given one. If we take $u=x\mp e^{i\arg(x_{j})}|x|e_{j}$, with
$(e_j)_k = \delta_{jk}$,
it can be easily proved that \[
P_{u}x=\pm e^{i\arg(x_{j})}|x|e_{j}\]
as required. The freedom on the sign in the expression defining the vector
$u$ can be used to make sure that the vector $u$ is non zero. 
The rest of the Householder tridiagonalization procedure follows exactly as
in the real case. Consider a skew-symmetric matrix of dimension N (even)
\begin{equation}
A=\left(\begin{array}{c|ccc}
0 & a_{12} & a_{13} & \ldots\\
\hline -a_{12}\\
-a_{13} &  & ^{(N-1)}A\\
\vdots\end{array}\right)\label{eq:A}
\end{equation}
The Householder transformation matrix is
\[
P_{1}=\left(\begin{array}{c|ccc}
1 & 0 & 0 & \ldots\\
\hline 0\\
0 &  & ^{(N-1)}P_{1}\\
\vdots\end{array}\right)
\]
where $^{(N-1)}P_{1}$ is built by using Eq. (\ref{eq:Householder})
and taking the vector $x$ (of dimension N-1) as $(a_{12},a_{13},\ldots)^{T}$.
The resulting transformed matrix is given by
\[
P_{1}AP_{1}^{T}=\left(\begin{array}{c|ccc}
0 & k_{1} & 0 & \ldots\\
\hline -k_{1}\\
0 &  & ^{(N-1)}\tilde{A}\\
\vdots\end{array}\right)
\]
where $k_{1}=\pm e^{i\arg(a_{12})}|x|$ and the matrix $^{(N-1)}\tilde{A}$
is skew-symmetric and given by
 $^{(N-1)}\tilde{A}={}^{(N-1)}P_{1}{}^{(N-1)}A{}^{(N-1)}P_{1}^{T}$.
Performing this procedure a total of N-2 times we end up with a tridiagonal
and skew-symmetric matrix
\begin{equation}
P_{N-2}\ldots P_{2}P_{1}AP_{1}^{T}P_{2}^{T}\ldots P_{N-2}^{T}=\left(\begin{array}{cccccc}
0 & k_{1} & 0 & 0 & \ldots & 0\\
-k_{1} & 0 & k_{2} & 0 & \ldots & 0\\
0 & -k_{2} & 0 & \ddots & \ldots & \vdots\\
 &  & \ddots & 0 & \ddots\\
0 &  &  & \ddots & 0 & k_{N-1}\\
\vdots &  & \vdots &  & -k_{N-1} & 0
\end{array}\right)\label{eq:AT}
\end{equation}
Using now a known property the Pfaffian (see \ref{sec:AppA}) we can deduce
from the above identity that $\det(P_{1})\ldots\det(P_{N-2})\pf(A)=\pf(A_{TR})$
where $A_{TR}$ is the triagonal and skew-symmetric matrix of the
right hand side of Eq. (\ref{eq:AT}). Taking into account that the
determinant of any Householder matrix is -1 and that $N$ is even,
we can express the Pfaffian of $A$ in terms of the pfaffian of the
tridiagonal $A_{TR}$ 
\[
\pf(A)=\pf(A_{TR})
\]
As will be shown below the Pfaffian of a tridiagonal skew-symmetric
matrix is simply given by $k_{1}k_{3}\ldots k_{N-1}=\prod_{i=1}^{N/2}k_{2i-1}$
(this result can also be obtained using the {}``minor expansion''
formula \cite{Caianello} ) and finally we obtain
\begin{equation}
\pf(A)=\prod_{i=1}^{N/2}k_{2i-1}.\label{eq:PfATri}
\end{equation}

In terms of numerical stability, the Householder transformation is
very robust and there is no need to consider any {}``pivoting''
strategy common to other methods.  However, the presence of the 
square root of $x$ and the argument $\arg(x_{j})$ of complex quantities 
prevents an easy implementation of the Householder tridiagonalization 
procedure for symbolic computation. For this purpose the second method
described in the next section is far easier to implement.

\subsection{Aitken's block diagonalization formula}

There is an alternative method for the calculation of the pfaffian,
which is also well suited for a symbolic implementation and that relies
on an expression for the pfaffian of a bipartite skew-symmetric matrix.
Let us start with a general skew-symmetric matrix $A$ (dimension
N, even) given by\begin{equation}
A=\left(\begin{array}{cc}
R & Q\\
-Q^{T} & S\end{array}\right)\label{eq:BP_S}\end{equation}
where $R$ and $S$ are square skew-symmetric matrices and $Q$ is
a general rectangular matrix (to account for the case where $R$ and
$S$ have different dimensions). Using Aitken's block diagonalization
formula (see \cite{Bunch.82} for an early use of the formula 
and \cite{Zhang.05} for a recent and thorough reference) for a bipartite
matrix we obtain \begin{eqnarray}
\left(\begin{array}{cc}
\mathbb{I} & 0\\
Q^{T}R^{-1} & \mathbb{I}\end{array}\right)\left(\begin{array}{cc}
R & Q\\
-Q^{T} & S\end{array}\right)\left(\begin{array}{cc}
\mathbb{I} & -R^{-1}Q\\
0 & \mathbb{I}\end{array}\right) & =\nonumber \\
\left(\begin{array}{cc}
R & 0\\
0 & S+Q^{T}R^{-1}Q\end{array}\right)\label{eq:PSPT}\end{eqnarray}
where the matrix $S+Q^{T}R^{-1}Q$ is referred to in the literature
as the Schur complement of the matrix $A$ (see, for instance, \cite{Zhang.05}).
For the special case of a skew-symmetric matrix $A$, the matrices
$R$ and $S$ are also skew-symmetric and the transformation of the
matrix $A$ is a congruence (i.e. the matrix acting on the left hand
side of $A$ is the transpose of the one acting on the right hand
side). Denoting
\begin{equation}
P_{1}=\left(\begin{array}{cc}
\mathbb{I} & 0\\
Q^{T}R^{-1} & \mathbb{I}\end{array}\right)\label{eq:P1}\end{equation}
Eq. (\ref{eq:PSPT}) becomes\[
P_{1}AP_{1}^{T}=\left(\begin{array}{cc}
R & 0\\
0 & S+Q^{T}R^{-1}Q\end{array}\right)\]
An equivalent expression involving $S^{-1}$ instead of $R^{-1}$
is easily obtained \[
P_{2}AP_{2}^{T}=\left(\begin{array}{cc}
R+QS^{-1}Q^{T} & 0\\
0 & S\end{array}\right)\]
with
\begin{equation}
P_{2}=\left(\begin{array}{cc}
\mathbb{I} & -QS^{-1}\\
0 & \mathbb{I}\end{array}\right)\label{eq:P2}\end{equation}
 By using the property $\textrm{pf}(P^{T}AP)=\textrm{det}(P)\textrm{pf}(A)$
(see \ref{sec:AppA}) and taking into account that $\det P_{1}=\det P_{2}$=1,
we come to
\begin{eqnarray}
\textrm{pf}(A) & = & \textrm{pf}(R)\textrm{pf}(S+Q^{T}R^{-1}Q)\label{eq:BP_PfS}\\
 & = & \textrm{pf}(R+QS^{-1}Q^{T})\textrm{pf}(S)\end{eqnarray}
Another nice property of the matrices $P_{1}$ and $P_{2}$ is that
their inverses can be obtained very easily \begin{equation}
P_{1}^{-1}=\left(\begin{array}{cc}
\mathbb{I} & 0\\
-Q^{T}R^{-1} & \mathbb{I}\end{array}\right)\label{eq:P1I}
\end{equation}
and
\begin{equation}
P_{2}^{-1}=\left(\begin{array}{cc}
\mathbb{I} & QS^{-1}\\
0 & \mathbb{I}\end{array}\right)\label{eq:P2I}
\end{equation}
These expressions of the inverses explicitly show that both $P_{1}$ and
$P_{2}$ are not orthogonal matrices. 

Let us now apply the above result to an arbitrary skew-symmetric matrix
of dimension $N=2M$ which is written in block form as \begin{equation}
A=\left(\begin{array}{ccc}
A^{(1)} & A_{N-1} & A_{N}\\
-A_{N-1}^{T} & 0 & a_{N-1,N}\\
-A_{N}^{T} & -a_{N-1,N} & 0\end{array}\right)\label{eq:S_Aitken}\end{equation}
where $A^{(1)}$ is a skew-symmetric square matrix of dimension $N-2=2(M-1)$
and $A_{N-1}$ and $A_{N}$ are column vectors $A_{N-1}=\{A_{i,N-1},\, i=1,N-2\}$
and $A_{N}=\{A_{i,N},\, i=1,N-2\}$ both of dimension $(N-2)\times1$.
In the language of Eq (\ref{eq:BP_S}) the matrix $R$ is the matrix
$A^{(1)}$, the matrix $Q$ is a rectangular matrix of dimension $2\times(N-2)$
made of the two column vectors, $A_{N-1}$ and $A_{N}$ and finally
the matrix $S$ is the $2\times2$ skew-symmetric matrix with matrix
element $S_{12}=a_{N-1,N}$. Using the ideas of Aitken's block diagonalization
formula, it is easy to shows that the matrix $\tilde{A}=D_{1}^{T}AD_{1}$
is in block diagonal form
\begin{equation}
\tilde{A}=\left(\begin{array}{ccc}
\mbox{\ensuremath{\tilde{A}}}^{(1)} & 0 & 0\\
0 & 0 & a_{N-1,N}\\
0 & -a_{N-1,N} & 0\end{array}\right)\label{eq:STilda}
\end{equation}
with a matrix $D_{1}$ of the form 
\begin{equation}
D_{1}=\left(\begin{array}{ccc}
\mathbb{I}_{N-2} & 0 & 0\\
X & 1 & 0\\
Y & 0 & 1\end{array}\right)\label{eq:D}
\end{equation}
where $\mathbb{I}_{N-2}$ stands for the identity matrix of dimension
$N-2$ and both $X$ and $Y$ are row vectors of dimension $1\times(N-2)$
and given by $X=-a_{N-1,N}^{-1}A_{N}^{T}$ and $Y=a_{N-1,N}^{-1}A_{N-1}^{T}$.
In the above equation \ref{eq:STilda}, the skew-symmetric matrix
$\tilde{A}^{(1)}$ is given by
\begin{equation}
\tilde{A}^{(1)}=A^{(1)}+A_{N}(a_{N-1,N})^{-1}A_{N-1}^{T}-A_{N-1}a_{N-1,N}^{-1}A_{N}^{T}
\label{eq:S1Trans}
\end{equation}
 Taking into account that $\det D_{1}=1$ then
$\pf(A)=\pf(\tilde{A})=a_{N-1.N}\pf(\tilde{A}^{(1)})$.
The algorithm can be applied recursively to $\tilde{A}^{(1)}$ to
obtain\\ $\pf(A)=a_{N-1,N}\tilde{a}_{N-3,N-2}^{(1)}\pf(\tilde{A}^{(2)})$
so as to reduce, after $M-1$ iterations, the computation of the pfaffian
to the product of the corresponding elements. 

This procedure can be easily implemented for a skew-symmetric tridiagonal
matrix, as the transformed matrices in Eq (\ref{eq:S1Trans}) coincide
with the original ones; for instance, $\tilde{A}^{(1)}=A^{(1)}$.
As a consequence, the pfaffian of a tridiagonal matrix is given by
\[
\textrm{pf }\left(\begin{array}{cccccc}
0 & d_{1} & 0 & 0 & \ldots & 0\\
-d_{1} & 0 & d_{2} & 0 & \ldots & 0\\
0 & -d_{2} & 0 & \ddots & \ldots & \vdots\\
0 & 0 & \ddots & 0 & \ddots & 0\\
\vdots & \vdots & \vdots & \ddots & 0 & d_{2N-1}\\
0 & 0 & \cdots & 0 & -d_{2N-1} & 0\end{array}\right)=d_{1}d_{3}\ldots 
d_{2N-1}=\prod_{i=1}^{N}d_{2i-1}\]

\subsubsection{Pivoting}

As a consequence of the division by matrix elements like $a_{N-1,N}$
in the first iteration, the numerical stability of the algorithm
requires the use of pivoting strategy in the implementation of the
method. Full pivoting amounts to search the whole matrix for the matrix
element with the largest modulus and exchange it with the required
matrix element. For instance, in the first iteration of the procedure,
the matrix element $a_{p,q}$ ($p<q)$ with the largest modulus is
searched for and exchanged with the matrix element $a_{N-1,N}$. In
this way we avoid dangerous divisions by small (or even zero) matrix
elements. We have to take into account that in the present case, the
exchange of both columns and rows is required to preserve the skew-symmetric
nature of the matrices involved. To carry out the exchange of rows
and columns we will use the exchange matrix $P(ij)$ that, when applied
to the right of an arbitrary matrix, exchanges columns $i$ and $j$.
The exchange matrix is given by the matrix elements 
\begin{eqnarray}
P(ij)_{kl} & = & \delta_{kl}-\delta_{i,l}\delta_{i,k}-\delta_{j,l}
\delta_{j,k}+\delta_{i,l}\delta_{j,k}+\delta_{j,l}\delta_{i,k}.\label{eq:Pij}
\end{eqnarray}
To exchange the corresponding rows we have to apply $P(ij)^{T}$ to
the left of the matrix (notice that $P(ij)$ is symmetric). With the
help of these matrices we can write the matrix after pivoting $a_{p,q}$
with $a_{N-1,N}$ (and $a_{q,p}$ with $a_{N,N-1})$ as\[
A_{P}=P^{T}(N-1,p)P^{T}(N,q)\, A\, P(N-1,p)P(N,q)\]
As a consequence of such exchange and taking into account that $\det P(ij)=-1$
we can conclude that the pfaffian of $A$ does not change by the pivoting
procedure. Finally we obtain 
$$
A=P(N,q)P(N-1,p)\, A_{P}\, P^{T}(N-1,p)P^{T}(N,q)
$$
$$
=P(N,q)P(N-1,p)D_{1}^{T\,-1}\tilde{A}_{P}D_{1}^{-1}P^{T}(N-1,p)P^{T}(N,q)
$$
where $\tilde{A}_{P}$ has the same structure as $\tilde{S}$ in Eq.
(\ref{eq:STilda}). As before,
$\pf(A)=\pf(\tilde{A}_{P})=\left(A_{P}\right)_{N-1,N}\pf(\tilde{A}_{P}^{(1)})$
and repeating recursively the whole procedure $M-1$ times we obtain
the pfaffian as the product of the corresponding matrix elements.

\subsubsection{Cholesky like decomposition of a skew-symmetric matrix}

Although it is not necessary in order to compute the pfaffian, it can
be useful to show that even with pivoting we can write the matrix
$A$ as
\begin{equation}
A=PL^{T}\tilde{A}LP\label{eq:Chol}
\end{equation}
where $P$ is the product of exchange matrices as in Eq (\ref{eq:Pij}),
$L$ is the product of matrices of the $D^{-1}$ type, Eq (\ref{eq:D}),
and therefore is a lower triangular matrix with ones in the main diagonal
and finally, $\tilde{A}$ is a skew-symmetric matrix in canonical
form, i.e. a block diagonal matrix with skew-symmetric, $2\times2$
blocks in the diagonal. This decomposition of a general skew-symmetric
matrix $A$ resembles the Cholesky decomposition of a general matrix
and can be useful in formal manipulations like, for instance, the
inversion of the matrix $A$. In order to show that Eq (\ref{eq:Chol})
holds the only required property is that, when applying the pivoting
procedure to $\tilde{A}_{P}^{(1)}$ the exchange matrices required
$P(N-2,s)P(N-3,r)$ have the property of preserving the structure
of the matrix $D_{1}$(and its inverse). For instance, \[
D_{1}^{T\,-1}P(N-2,s)P(N-3,r)=P(N-2,s)P(N-3,r)\tilde{D}_{1}^{T\,-1}\]
with $\tilde{D}_{1}^{T\,-1}$ a matrix that is obtained from $D_{1}^{T\,-1}$
by exchanging rows $N-2$ and $s$ and rows $N-3$ and $r$ and therefore
has the same upper triangular structure with ones in the diagonal
as the original matrix $D_{1}^{T\,-1}$ . Using this property we can
move all the exchange matrices to the right (or to the left) and the
remaining matrix will be the product of triangular matrices (lower
for products involving $D^{-1}$) with ones in the diagonal.

As mentioned earlier, Aitken's method is better adapted to symbolic 
evaluations.  However, one must take care that in each step of the 
process some specific matrix elements are non-zero. 

\section{Fortran implementation}

The implementation of the algorithms considered in this paper in a
high level computer language is straightforward. However, specific
code in FORTRAN (both 77 and 90, real and complex arithmetic) is provided
along with this paper. The algorithms are easy to follow and the comments
included in the code are useful guides. Just a few comments are in
order: to implement the tridiagonalization procedure in Fortran, it
is advantageous to use the BLAS package \cite{BLAS} to perform the
required matrix by vector multiplication and rank two update. Unfortunately
there are no equivalent in the skew-symmetric case of the routines
SYM (to multiply a symmetric matrix by a vector) or SYR2 (to perform
a symmetric rank two update) but the general procedures GEMV and GERU
can be used instead. On output, both the pfaffian of the matrix and
the set of vectors required to bring it to tridiagonal form are returned.
For the implementation of the method based on Aitken's block diagonalization
formula a pivoting strategy is required. We have used full pivoting
in our implementation due to its robustness. The routines provided
only require the upper part of the skew-symmetric matrix. The lower
part is destroyed and replaced with the tridiagonal transformation
matrix that brings the skew-symmetric matrix to canonical form upon
congruence. An integer vector is also returned to reconstruct the
required exchange of rows and columns.

Perhaps the best test to check the validity of the two implementations
is to compute the pfaffian of a skew-symmetric matrix using both procedures
in order to compare the output. If it is the same up to a given accuracy
then it is very likely that the two implementations are correct. We have 
writen a test program
(also included in the distribution) that generates skew-symmetric
matrices of given dimension with random entries and compute the pfaffian
using both techniques. In our tests the pfaffians computed both ways
coincide up to one part in $10^{10}$ with dimensions of the matrices of
one thousand. This result also supports the adequacy of the implementation
in terms of numerical stability.
Another possibility to test the numerical implementation
is to use the analytical formula given in \ref{sec:PfaffTestMatrix}
for a specific kind of $8\times8$ matrices. A test program implementing
this approach has also been included in the distribution.

To finish this section we will briefly comment on the timing of the
FORTRAN numerical implementations mentioned. In a modern personal
computer under Linux the computation of the pfaffian of a $100\times100$
matrix takes a few milliseconds in both implementations and the timing
scales roughly as the cube of the dimension of the matrix in such a way
that for matrices of a $1000\times1000$ dimension the time is of the order
of a few seconds.

\section{A simple Python implementation}

We provide here a simple implementation of the tridiagonal reduction method
(see \cite{Bunch.82} and \cite{Aasen.71})
in Python, which may be useful for testing purposes.  It is similar to
the Householder, but it only use simple row and column operations 
that have determinants of unity.  
The code is:
\begin{verbatim}
from numpy import *
def pfaff_py(m) :
  mat=copy(m)
  ndim = shape(mat)[0]
  t1=1.0
  for j in range(ndim/2) :
    t1 *= mat[0,1]
    print 't1', t1
    if j <ndim/2-1 :
      ndimr=shape(mat)[0]
      for i in range(2,ndimr) :
        if mat[0,1] != 0.0 :
          tv=mat[1,:]*mat[i,0]/mat[1,0]
          mat[i,: ] -= tv
          tv=mat[:,1]*mat[0,i]/mat[0,1]
          mat[:,i ] -= tv
        else :
          print 'need to pivot'
          raise Exception
      mat=mat[2:,2:] 
  return t1
\end{verbatim}
The user should be cautioned that the algorithm is not
guaranteed to be stable without an additional pivot step. Also, the 
matrix is assumed to have been constructed with the
{\tt array} function in the Numpy library.

\section{A simple Mathematica implementation}
We also provide a simple Mathematica implementation of the method based on 
Aitken's block diagonalization formula. As mentioned above, this method
requires pivoting to avoid divisions by small (or zero) numbers. 
In the symbolic implementation, this issue is solved by replacing
the denominator by a variable (OO on the implementation below) 
in case it is zero and an additional limit when the variable tends to
zero is performed at the end.
The two Mathematica modules required are:
\begin{verbatim}
Aitken[M_,n_,OO_]:=
Module[{MM=M,i,j,p},
If[MM[[n-1,n]]==0,MM[[n-1,n]]=OO;MM[[n,n-1]]=-OO];
p=MM[[n-1,n]];
For[i=1,i<=n-2,i++,
   For[j=1,j<=n-2,j++,  
      MM[[i,j]]=M[[i,j]]+(MM[[i,n]]*MM[[j,n-1]]-MM[[j,n]]*MM[[i,n-1]])/p
      ]
   ];
 MM];

pfaffian[S_]:=
Module[{T=S,n,p},
n=Length[T]/2;
If[T[[2*n-1,2*n]]==0,T[[2*n-1,2*n]]=OO;T[[2*n,2*n-1]]=-OO];
For[n=Length[T]/2;p=T[[2*n-1,2*n]],n>1,n--,
   T=Aitken[T,2*n,OO];
   p=p*T[[2*(n-1)-1,2*(n-1)]]
   ];
Limit[p,OO->0]];
\end{verbatim}

\section{Conclusions}

The issue of how to compute both numerically and symbolically the
pfaffian of a skew-symmetric matrix has been addressed using two different
approaches. Numerical stability issues are discussed and methods to
assure the desired accuracy are fully incorporated. A collection of
subroutines and test programs in FORTRAN (both 77 and 90, double precision
and complex) are provided. A few comments on the implementation of
the algorithms in Mathematica and Python are also given. 

\section{Acknowledgements}

We acknowledge K. Roche for a careful reading of the manuscript and 
several suggestions. This work was supported by MICINN (Spain) under 
research grants FPA2009-08958, and FIS2009-07277, as well as by 
Consolider-Ingenio 2010 Programs CPAN CSD2007-00042 and MULTIDARK 
CSD2009-00064.

\appendix

\section{Definition and basic properties of the pfaffian\label{sec:AppA}}

The pfaffian of a skew-symmetric matrix $R$ of dimension $2N$ and
with matrix elements $r_{ij}$ is defined as 
\[
\textrm{pf}(R)=\frac{1}{2^{n}}\frac{1}{n!}\sum_{\textrm{Perm}}\epsilon(P)r_{i_{1}i_{2}}r_{i_{3}i_{4}}r_{i_{5}i_{6}}\ldots r_{2n-1,2n}\]
where the sum extends to all possible permutations of 
$i_{1},\ldots,i_{2n}$ and $\epsilon(P)$ is the parity of the permutation. For matrices of odd
dimension the pfaffian is by definition equal to zero. As an example,
the pfaffian of a $2\times2$ matrix $R$ is $\textrm{pf}(R)=r_{12}$
and for a $4\times4$ one $\textrm{pf}(R)=r_{12}r_{34}-r_{13}r_{24}+r_{14}r_{23}$.
Useful properties of the pfaffian are

\begin{equation}
\textrm{pf}(P^{T}RP)=\textrm{det}(P)\textrm{pf}(R),\label{eq:PTRP}\end{equation}

\[
\text{\textrm{pf}}\left(\begin{array}{cc}
0 & R\\
-R^{T} & 0\end{array}\right)=(-1)^{N(N-1)/2}\det(R)\]
where the matrix $R$ is $N\times N$ and \[
\text{\textrm{pf}}\left(\begin{array}{cc}
R_{1} & 0\\
0 & R_{2}\end{array}\right)=\text{\textrm{pf}}(R_{1})\text{\textrm{pf}}(R_{2})\]
where $R_{1}$ and $R_{2}$ are skew-symmetric matrices. The matrices may
be defined on the real or on the complex number fields.

\section{Pfaffian of a test matrix\label{sec:PfaffTestMatrix}}

In this appendix we give the expression of the pfaffian of a test
matrix which is big enough as not to be trivial but on the other
hand is small enough as to render the explicit expression of the pfaffian
manageable. The expression given below can be used to check both numerical
and symbolic implementations of the pfaffian. 

Consider the two general skew-symmetric matrices of dimension 4\[
M=\left(\begin{array}{cccc}
0 & f_{1} & m_{11} & m_{12}\\
-f_{1} & 0 & m_{21} & m_{22}\\
-m_{11} & -m_{21} & 0 & f_{2}\\
-m_{12} & -m_{22} & -f_{2} & 0\end{array}\right)\]
and \[
N=\left(\begin{array}{cccc}
0 & g_{1} & n_{11} & n_{12}\\
-g_{1} & 0 & n_{21} & n_{22}\\
-n_{11} & -n_{21} & 0 & g_{2}\\
-n_{12} & -n_{22} & -g_{2} & 0\end{array}\right)\]
where the matrix elements can be complex numbers. With these two matrices
and the identity $4\times4$ matrix we build the skew-symmetric matrix\[
S=\left(\begin{array}{cc}
N & -\mathbb{I}\\
\mathbb{I} & -M^{*}\end{array}\right)\]
of dimension $8\times8$ (see Ref \cite{Robledo.09} for the physical
context of this matrix). It is relatively easy to compute its pfaffian
$$
\pf[S]=1+f_{1}^{*}g_{1}+f_{2}^{*}g_{2}+
m_{11}^{*}n_{11}+m_{22}^{*}n_{22}+m_{12}^{*}n_{12}+m_{21}^{*}n_{21}+
$$
$$
+(f_{1}^{*}f_{2}^{*}-m_{11}^{*}m_{22}^{*}+
m_{12}^{*}m_{21}^{*})(g_{1}g_{2}-n_{11}n_{22}+n_{12}n_{21})
$$

\end{document}